\documentclass[final,3p,times,twocolumn]{elsarticle}
\usepackage{bbm}
\usepackage{amsmath}
\usepackage{hyperref}
\usepackage[linesnumbered,ruled]{algorithm2e}
\usepackage{url}

\journal{Journal of \LaTeX\ Templates}
\begin{document}
\begin{frontmatter}
\title{A Teacher-Student Framework with Fourier Transform Augmentation for COVID-19 Infection Segmentation in CT Images}

\author[mymainaddress]{Han Chen}
\author[mymainaddress]{Yifan Jiang}
\author[mymainaddress]{Hanseok Ko\corref{mycorrespondingauthor}}
\address[mymainaddress]{School of Electrical Engineering, Korea University, Seoul, South Korea}
\cortext[mycorrespondingauthor]{Corresponding author}
\ead{hsko@korea.ac.kr}
\author[mysecondaryaddress]{Murray Loew}
\address[mysecondaryaddress]{Biomedical Engineering, George Washington University, Washington D.C., USA}

\begin{abstract}
Automatic segmentation of infected regions in computed tomography (CT) images is necessary for the initial diagnosis of COVID-19. Deep-learning-based methods have the potential to automate this task but require a large amount of data with pixel-level annotations. Training a deep network with annotated lung cancer CT images, which are easier to obtain, can alleviate this problem to some extent. However, this approach may suffer from a reduction in performance when applied to unseen COVID-19 images during the testing phase, caused by the difference in the image intensity and object region distribution between the training set and test set. In this paper, we proposed a novel unsupervised method for COVID-19 infection segmentation that aims to learn the domain-invariant features from lung cancer and COVID-19 images to improve the generalization ability of the segmentation network for use with COVID-19 CT images. First, to address the intensity difference, we proposed a novel data augmentation module based on Fourier Transform, which transfers the annotated lung cancer data into the style of COVID-19 image. Secondly, to reduce the distribution difference, we designed a teacher-student network to learn rotation-invariant features for segmentation. The experiments demonstrated that even without getting access to the annotations of the COVID-19 CT images during the training phase, the proposed network can achieve a state-of-the-art segmentation performance on COVID-19 infection.
\end{abstract}

\end{frontmatter}

\section{Introduction}
The pandemic caused by the novel coronavirus disease (COVID-19) that emerged at the start of 2020 has had significant worldwide medical and economic impacts \cite{wang2020novel,shi2020review}. A recent report by the World Health Organization found that there had been more than 198 million confirmed cases and more than 5 million deaths globally by January 2022 \cite{covid19}. To diagnose COVID-19, real-time reverse transcription-polymerase chain reaction (RT–PCR) tests and radiological imaging techniques, such as computed tomography (CT) and X-rays, are widely used. In particular, CT imaging plays a critical role in the early diagnosis and evaluation of COVID-19 \cite{lei2020ct,li2020artificial}, with the segmentation of infected regions in CT scans providing essential information that can be used in the quantitative assessment of the progression of the disease \cite{huang2020serial,cao2020longitudinal}. 

\begin{figure}[!t]
\centering
\includegraphics[width=8cm, height=3.5cm]{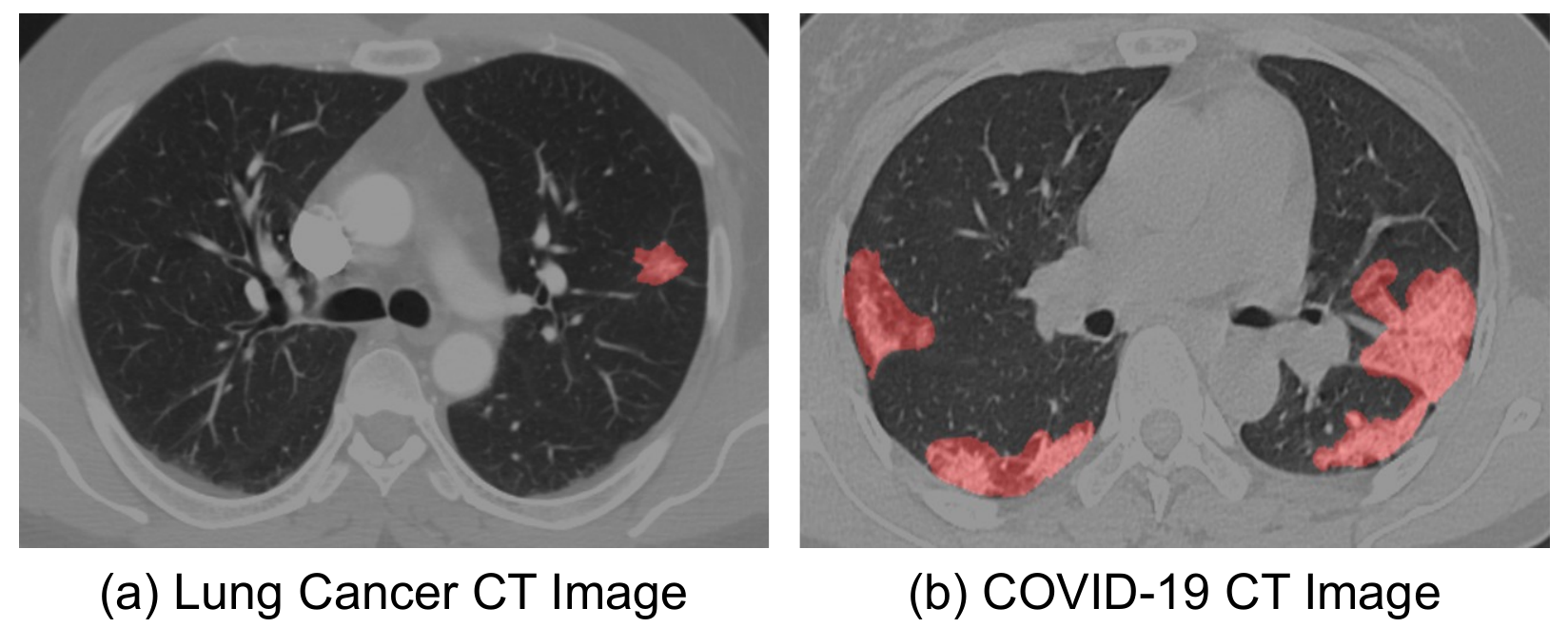}
\caption{Comparison of the pulmonary nodule and COVID-19 infection in CT images. (a) shows a typical nodule of lung cancer and (b) illustrates the infection of a COVID-19 confirmed case. The nodule and infection are marked with red.}
\label{fig:comparision} 
\end{figure}

Overall, segmentation approaches can be split into three groups: (1) Manual-based segmentation is defined as the delineation of the borders of anatomical regions that are conducted by experts (e.g., radiologists, pathologists) \cite{bouchareb2021artificial}; (2) Conventional methods-based segmentation is described as the assignment of labels to pixels or voxels by matching the prior known object model to the image data according to the radiomics features (hand-crated or explicit features) \cite{sharafudeen2022detecting}; (3) Deep-learning-based segmentation as developed for automated feature extraction \cite{han2019hybrid,jin2021memory,fusco2021artificial}. Segmentation of the lungs in COVID-19 cases involves outlining the contours of the anatomical structures of the lung or infection with computer-assisted techniques in CT or X-ray images. Due to the high variety in shape, size, boundary, type, and manifestation of regions of interest (ROI), the conventional algorithms failed to segment the infected region properly \cite{bouchareb2021artificial}. Some deep-learning-based automatic segmentation methods have been proposed for use in COVID-19 segmentation and achieved impressive results \cite{gao2021dual,laradji2021weakly,paluru2021anam}. 

However, current COVID-19 infection segmentation approaches rely on large datasets with pixel-level annotations, which is time-consuming and laborious to collect due to privacy issues and the lack of experts \cite{adler2012sharing,sharma2019preserving,jiang2020covid}. In contrast, as one of the most common cancers worldwide, the research on lung cancer is more widely, and the labeled lung cancer CT data is more accessible \cite{xu2019deep}. Therefore, it is possible to utilize publicly available lung cancer databases to promote the performance of a deep network for the segmentation of COVID-19 infection. For example, Jin et al. \cite{jin2020development} developed a system using a large dataset from the Lung Image Database Consortium and Image Database Resource Initiative (LIDC-IDRI) \cite{armato2011lung} to achieve multi-class classification diagnoses. Wang et al. \cite{wang2021does} revealed that transferring knowledge of lung cancer datasets can boost the segmentation performance for COVID-19 infection. They proved that a robust pre-training model can be obtained by training a segmentation model with lung cancer data with pulmonary nodule, which benefits subsequent supervised learning on a small COVID-19 dataset.

Even though COVID-19 infection and pulmonary nodule exhibit similar manifestations, few studies have directly used lung cancer datasets to train the deep network for COVID-19 infection segmentation. Models only trained with lung cancer CT images will suffer a significant performance drop when tested on COVID-19 CT images due to the domain shift between the two datasets. The difference between pulmonary nodules and COVID-19 infection in CT scans is presented in Fig.\ref{fig:comparision} and can be categorized as follows: (1) In terms of distribution, COVID-19 presents as a bilateral, patchy infection, while early-stage lung cancer is unilateral and oval in shape \cite{zhang2020covid}; (2) While there is also a clear difference in intensity due to the use of different scanners, scanning protocols, and subject populations \cite{guan2021domain}.

In this paper, we consider COVID-19 infection segmentation in the context of the wide availability of lung cancer CT images with annotations, the limited availability of unlabeled COVID-19 CT images, and the difference between these two domains. Our motivation is that the features learned from pulmonary nodules in lung cancer CT can be used for the segmentation of COVID-19 infection \cite{wang2021does}, making it possible to construct an unsupervised COVID-19 infection segmentation method. For our segmentation network, we designed a novel COVID-19 style guided Fourier Transform-based data augmentation method (CGFT-DA) and a training scheme to align these two datasets. In order to address the intensity difference, we transferred the lung cancer CT images into the style of COVID-19 CT images with the designed CGFT-DA module, which replaces the low-level frequency information of the lung cancer CT images with that of COVID-19 CT images. The output of the CGFT-DA module will keep the same semantic information with the lung cancer image but in the COVID-19 image style. Since the lung cancer CT images are labeled at pixel-level, the transferred images and corresponding annotations can be used to train the base segmentation network. To overcome the distribution difference, we introduced a teacher-student learning paradigm to achieve robust features learning. We treated our base segmentation network as the student network and introduced another teacher network, and then imposed the same elastic transformation on the input to the student network and the output of the teacher network, respectively. The output predictions of these two networks are forced to be consistent. We validated the effectiveness of our proposed method with public COVID-19 CT images. Experimentally, it outperformed various competing state-of-the-art approaches. 

The contributions of our work can be summarized as follows:
\begin{itemize}
\item We proposed a novel unsupervised COVID-19 infection segmentation method. To the best of our knowledge, our method is the first attempt to utilize the lung cancer data for the COVID-19 infection segmentation task.

\item We proposed an effective data augmentation module to overcome the intensity difference between the lung cancer CT data and COVID-19 CT data via the Fourier Transform and its inverse.

\item We built a framework that collaborates with a teacher-student network based on transformation-consistency learning, which alleviates the distribution difference and enables the network to learn the robust features of COVID-19 infection in CT images.

\item The experimental results showed that our proposed method can achieve state-of-the-art performance for COVID-19 infection segmentation. We also provided a comprehensive analysis of our method.
\end{itemize}

\section{Related Works}

\paragraph{COVID-19 Infection Segmentation} Deep-learning-based segmentation methods have played an essential role in diagnosing COVID-19 \cite{chen2021unsupervised,pathak2020deep,ozkaya2020coronavirus,verma2022covxmlc,bhattacharyya2022deep}. Some studies explored to construct new networks suitable for COVID-19 CT segmentation task \cite{ouyang2020dual,oulefki2021automatic,punn2022chs}. For example, Ouyang et al. \cite{ouyang2020dual} developed an online module with a 3D convolutional network (CNN) for better localization of COVID-19 infected areas. Other works utilized attention mechanism to learn rich contextual relationships for better feature representations \cite{zhou2020automatic,karthik2022contour,hu2022deep}. Zhou et al. \cite{zhou2020automatic} proposed to use spatial and channel attention to improve the representation ability of the segmentation network. To address the issue of scarcity of well-labeled data, solutions based on weak supervised, semi-supervised learning and multi-task learning were proposed in the literature \cite{fan2020inf,liu2022weakly,malhotra2022multi}. Fan et al. \cite{fan2020inf} presented a semi-supervised segmentation method based on a random selection propagation strategy, which required only a few labeled images and primarily utilizes unlabeled data. However, current techniques still rely more or less on labeled COVID-19 data and cannot handle the more challenging unsupervised situation. Meanwhile, few studies consider using lung cancer data which is easier to access, to boost the segmentation performance of COVID-19 infection.

\paragraph{Semi-supervised Learning for Medical Segmentation} Deep-learning-based methods usually require a large amount of labeled data, however, it is expensive and time-consuming to annotate data for medical segmentation tasks. Semi-supervised learning gained significant attention as a way to utilize the unlabeled data \cite{jiang2021few}. Some studies are based on self-training, which consists of pre-training the segmentation network using unlabeled medical images followed by fine-tuning the pre-trained model using the target labeled dataset \cite{qiao2018deep,yang2020fda,li2022artificial}. Other works are based on pseudo annotations, they first assigned pseudo annotations to unlabeled data and then trained the segmentation model using both the labeled and pseudo labeled data \cite{luo2021semi,chen2021semi,ouali2020semi}. For example, Luo et al. \cite{luo2021semi} proposed to utilize cross training and take the prediction of a network as pseudo-label for other networks. Another strategy is training a model with both labeled and unlabeled data, where the unlabeled data is used to generate a supervision signal through an unsupervised loss function \cite{zhang2017deep,vu2019advent,verma2022interpolation}. Zhang et al. \cite{zhang2017deep} proposed an adversarial network for biomedical image segmentation, which is composed of an evaluation network to assess the segmentation performance and encourage the network to attain consistently good results on both annotated and unannotated images.

\paragraph{Teacher-Student Learning} The concept of teacher-student learning is based on distilling the knowledge of an ensemble of deep networks into a single network \cite{tarvainen2017mean,bucilua2006model}. It employs unlabeled data to produce consistent predictions under perturbations and has been explored in segmentation tasks \cite{yu2019uncertainty,hang2020local,jin2021trseg}. Yu et al. \cite{yu2019uncertainty} proposed training a teacher-student model in which the teacher model produced noisy labels and corresponding uncertainty maps, while the student model was trained using the labels from the teacher model in consideration of label uncertainty. Xu et al. \cite{xu2019self} proposed a self-ensembling attention network consisting of a student model that acts as a base network and a teacher model serving as the ensembling network, the former learns from the output of the latter and tends to produce accurate predictions. Choi et al. \cite{choi2019self} utilized GAN to transfer the labeled image into the same style as the unlabeled image and then introduced a teacher network to enhance its performance in the segmentation of the unlabeled data.

To compensate for the data shortage issue for the COVID-19 infection segmentation task, training one semi-supervised network with large-scale lung cancer CT data and unlabeled COVID-19 CT data could be a possible solution. However, the current semi-supervised techniques mainly focus on utilizing the unlabeled data, which shares similar distribution with the labeled data due to coming from the same source. Therefore they cannot address the more challenging situation of significant differences between these two datasets. To solve the above problem, we introduced our method. Specifically, we proposed to transfer the lung cancer data into the style of COVID-19 CT images for data augmentation to overcome the intensity difference. By constructing a teacher-student network with consistency learning and entropy minimization, our network can learn the features that are robust to elastic transformations, at the same time, the distribution difference gets alleviated effectively.

\begin{figure*}[!t]
\centering
\includegraphics[width=16cm, height=5.5cm]{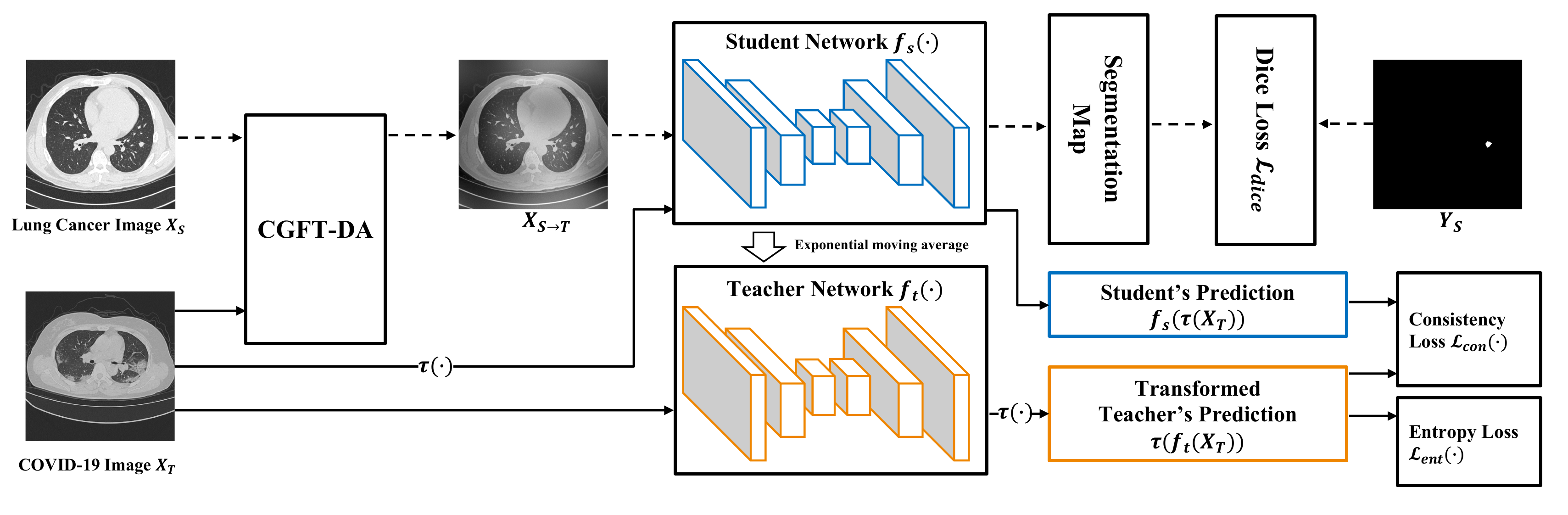}
\caption{Overview of the proposed method. It consists of a CGFT-DA module, a student network, and a teacher network. The dashed lines and solid lines represent the data flow for the lung cancer images and COVID-19 images, respectively. The student network is trained by the weighted combination of dice loss, consistency loss, and entropy loss. The weights of the teacher network are the exponential moving average of that of the student network. $\tau(\cdot)$ represents an elastic transformation operation with the same parameters in every iteration.}
\label{fig:overall}       
\end{figure*}

\section{Proposed Method}

In this section, we provide an overview of our proposed method. We then illustrate our Fourier Transform-based data augmentation method. Finally, we describe our teacher-student training strategy and optimization objection.

\subsection{Overview}
Fig.\ref{fig:overall} overviews our unsupervised COVID-19 infection segmentation network. The labeled lung cancer images and unlabeled COVID-19 CT images are denoted as $\{{X}_S, {Y}_S\}$$\sim$$ D_{S}$ and $\{{X}_T\}$$\sim$$ D_{T}$, respectively. We first send ${X}_S$ and ${X}_T$ to CGFT-DA module and obtain $\{X_{S\rightarrow{T}},Y_S\}$$\sim$$D_{S\rightarrow{T}}$ which retains the semantic information for lung cancer but in COVID-19 style. A teacher and a student network are employed in our framework, with the former acting as an ensemble network and the latter acting as the base segmentation network. We send ${X}_{S\rightarrow{T}}$ into the student network and obtain the pixel-wise segmentation prediction. ${X}_T$ will be sent to both the student and teacher networks, we then apply the same elastic transformation to the input of the student network and to the output of the teacher network, the two-stream outputs are aligned by the consistency loss. Back-propagation only occurs for the student network, and the weights of the teacher network are updated using the exponential moving average (EMA) of the student network.

\begin{figure}[t]
\centering
\includegraphics[width=8cm, height=5cm]{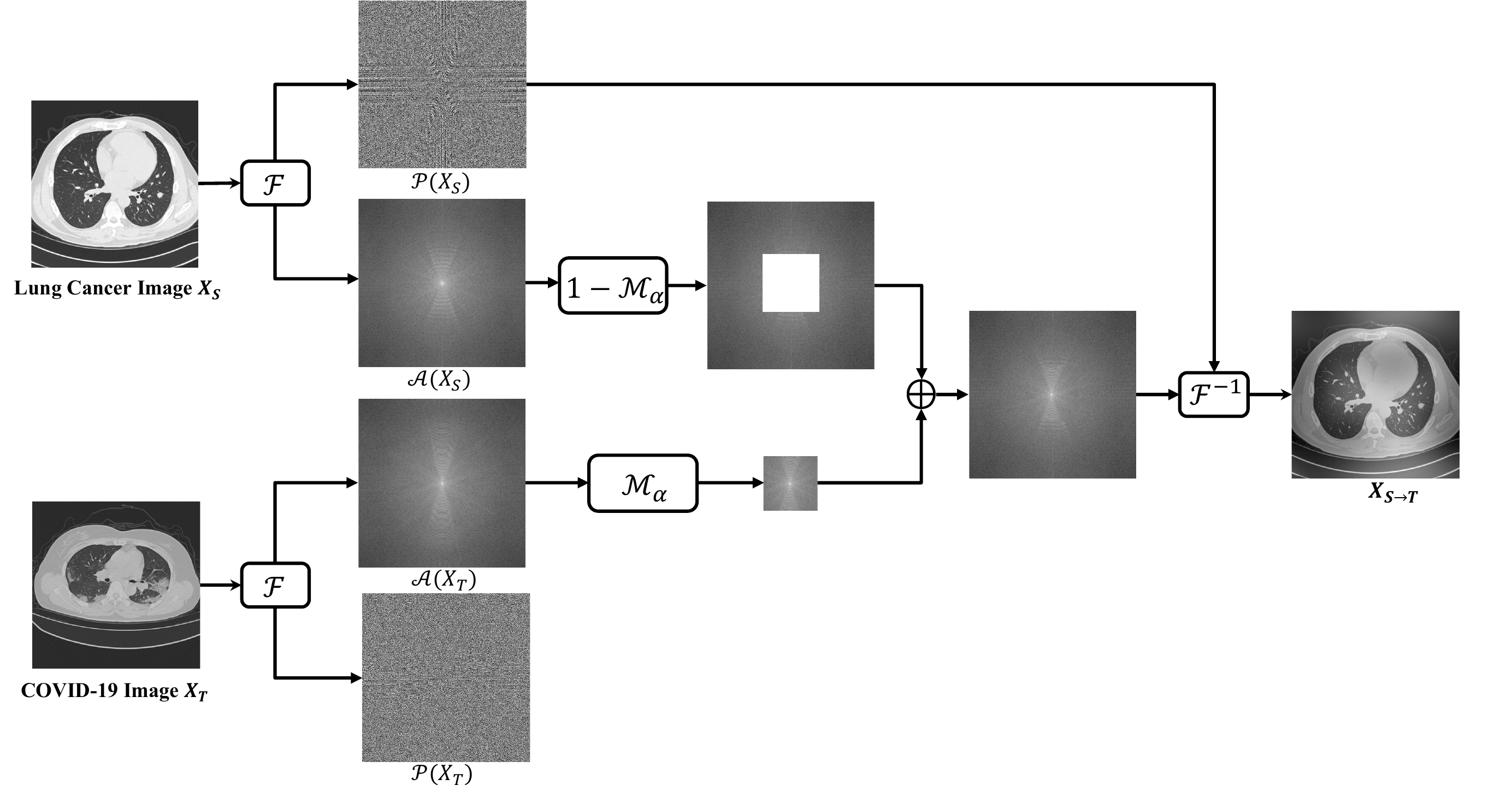}
\caption{The overview of the CGFT-DA module. The Fourier Transform is first conducted on the two inputs, and part of the amplitude spectrum from the lung cancer image is replaced with that from the COVID-19 image. After inverse Fourier Transform, the output image will keep the semantic information of the lung cancer image but with COVID-19 style.}
\label{fig:CGFT-DA}       
\end{figure}

\subsection{COVID-19 style guided and Fourier Transform-based data augmentation}
To address the intensity difference between the lung cancer and COVID-19 CT images, we proposed a COVID-19 style guided Fourier Transform-based data augmentation module (CGFT-DA), as shown in Fig.\ref{fig:CGFT-DA}. We employ the Fourier Transform and replace the low-frequency spectrum information of the lung cancer images with that of the COVID-19 images. In this way, we make it possible to disentangle the low-level distribution (i.e., the intensity information) from the high-level semantic information (i.e., the object content) of an image and transfer the former to another image.

Specifically, given an image $X\in\mathbb{R}^{H\times{W}\times{C}}$ ($C$=$1$ for a single-channel input), we can calculate its Fourier Transform using the FFT algorithm as,

\begin{equation}
\mathcal{F}(X)(u,v)=\sum\limits_{h=0}^{H-1}\sum\limits_{w=0}^{W-1}X(h,w)e^{-j2\pi(\frac{h}{H}u+\frac{w}{W}v)}
\label{eq1}
\end{equation}

We further decompose $\mathcal{F}(X)$ into an amplitude spectrum $\mathcal{A}(X)\in\mathbb{R}^{H\times{W}\times{C}}$ and a phase spectrum $\mathcal{P}(X)\in\mathbb{R}^{H\times{W}\times{C}}$, which represent the intensity distribution and semantic content of the image, respectively. We then denote a binary mask $\mathcal{M}_\alpha=\mathbbm{1}_{(h,w)\in[-\alpha{H}:\alpha{H},-\alpha{W}:\alpha{W}]}$, which controls the scale of the amplitude spectrum to be replaced. Given $\mathcal{A}(X_S)$ from the lung cancer image and $\mathcal{A}(X_T)$ from the COVID-19 image, we apply the mask $\mathcal{M}_\alpha$ to them and generate a new amplitude spectrum:

\begin{equation}
\mathcal{A}(X_{S\rightarrow{T}})=\mathcal{M}_\alpha\mathcal{A}(X_T)+(1-\mathcal{M}_\alpha)\mathcal{A}(X_S)
\label{eq2}
\end{equation}

After obtaining the transferred amplitude spectrum $\mathcal{A}(X_{S\rightarrow{T}})$, we combine it with the phase spectrum $\mathcal{P}(X_S)$ for lung cancer image $X_S$ and conduct the inverse Fourier Transform $\mathcal{F}^{-1}$ to generate transferred image $X_{S\rightarrow{T}}$.

\begin{equation}
X_{S\rightarrow{T}}=\mathcal{F}^{-1}(\mathcal{A}(X_{S\rightarrow{T}}), \mathcal{P}(X_S))
\label{eq3}
\end{equation}

\noindent where the semantic content of $X_{S\rightarrow{T}}$ is the same as $X_S$, but follows the intensity distribution of $X_T$. In this way, we obtain the transferred lung cancer data $\{X_{S\rightarrow{T}},Y_S\}$ with a similar intensity to that of COVID-19 images; this transferred data is utilized for further training and can effectively alleviate the intensity distribution between the two datasets.

\subsection{Teacher-student network and entropy minimization}
Even though the transferred data $X_{S\rightarrow{T}}$ have a similar style to $X_T$, a difference in distribution remains. To account for this distribution difference between the pulmonary nodules and COVID-19 infection, we built a teacher-student network with a transformation-consistency learning strategy. The teacher and student networks follow the same U-Net architecture \cite{ronneberger2015u} but are updated in different ways.

Specifically, the transferred lung cancer images $X_{S\rightarrow{T}}$ is only sent to the student network to output the pixel-wise segmentation result $\hat{Y}_S$, which is then utilized to calculate the following dice loss \cite{drozdzal2016importance}:

\begin{equation}
\mathcal{L}_{dice} = \frac{2*\sum_{n=1}^N{p_n}{g_n}}{\sum_{n=1}^N{p_n}+\sum_{n=1}^N{g_n}+\epsilon}
\label{eq4}
\end{equation}

\noindent where $g_n$ and $p_n$ represent the ground truth for the input and the predicted probabilistic map, respectively, with a background class probability of $1$$-$$p_n$. The term $\epsilon$ is used to avoid being divided by zero.

Updating the base student network with the dice loss in Equation (\ref{eq4}) leads to good segmentation performance for lung cancer images. However, when testing on the COVID-19 images, the performance will be significantly lower due to the distribution difference. To resolve this problem, we introduced a teacher network and utilize the unlabeled COVID-19 CT images to guide the student network to learn the robust features. As shown in Fig.\ref{fig:overall}, we first impose an elastic transformation $\tau(\cdot)$ on $X_T$, then the student network takes the transformed $\tau(X_T)$ as input and produces the predicted segmentation map. For the teacher network, we apply the same transformation $\tau(\cdot)$ on its prediction map. We then define the consistency loss to force the predictions for COVID-19 infection from these two networks to be consistent:

\begin{equation}
\mathcal{L}_{con}=\sum\limits_{i\in{\mathcal{I}}}{\|{{f_s(\tau(X_T))_{i}}-{\tau(f_t(X_T))}_{i}}\|}^2 
\label{eq5}
\end{equation}

\IncMargin{1em}
\label{alg:training}
\begin{algorithm}[t]\SetKwData{Left}{left}\SetKwData{This}{this}\SetKwData{Up}{up} \SetKwFunction{Union}{Union}\SetKwFunction{FindCompress}{FindCompress} \SetKwInOut{Input}{Input}\SetKwInOut{Output}{Output}\SetKwInOut{Parameters}{Parameters}
	
	\Input{CGFT-DA processed dataset $D_{S\rightarrow{T}}$ \\ COVID-19 dataset $D_T$}
	\Output{Trained parameters $\theta^*$}
	\Parameters{Two coefficients $\lambda$ and $\beta$ \\ An elastic transform function $\tau$ \\ An model forward function $f$}
	 \BlankLine 
	 
	 \emph{Initialize network parameters $\theta^*$ randomly}\; 
	 \For{$j=1$ to $N$}{ 
	 	\emph{$X_{S\rightarrow{T}},Y_S \sim D_{S\rightarrow{T}}$}\;
	 	\emph{$X_T \sim D_T$}\;
	 	\emph{$\hat{{Y_{S}}} \leftarrow f_{S}(X_{S\rightarrow{T}})$}\;
	 
	 	\emph{Compute Dice Loss $\mathcal{L}_{dice} \leftarrow \frac{2*\sum_{n=1}^N{p_n}{g_n}}{\sum_{n=1}^N{p_n}+\sum_{n=1}^N{g_n}+\epsilon}$}\;

	 	\emph{Compute Consistency Loss $\mathcal{L}_{con} \leftarrow \sum\limits_{i\in{\mathcal{I}}}{\|{{f_s(\tau(X_T))_{i}}-{\tau(f_t(X_T))}_{i}}\|}^2$}\;

	 	\emph{Compute Entropy Loss $\mathcal{L}_{ent} \leftarrow {-}\sum\limits_{i\in{\mathcal{I}}} f_t(X_T)_{i}log{(f_t(X_T))_{i}}$}\;

	 	\emph{$\mathcal{L}_{student}=\mathcal{L}_{dice}+\lambda\mathcal{L}_{con}+\mathcal{L}_{ent}$}\;
	 	
	 	\emph{Update $\theta_s$ by backpropagation}\;
	 	\emph{Update $\theta_t$ using EMA $\theta_{t,i}=\beta\theta_{t, j-1}+(1-\beta)\theta_{s, j}$}
 	 } 
 	 	  \caption{Model Training}
 	 	  \label{psedo_code} 
 	     \end{algorithm}
 \DecMargin{1em} 

To more firmly bridge the gap between the lung cancer images and the COVID-19 images, we introduced entropy minimization to force the model's decision boundary toward a high prediction certainty for the teacher network's prediction from the COVID-19 input. Given a COVID-19 CT image $X_T$, the entropy loss is calculated as follows:

\begin{equation}
\mathcal{L}_{ent}={-}\sum\limits_{i\in{\mathcal{I}}} f_t(X_T)_{i}log{(f_t(X_T))_{i}}
\label{eq6}
\end{equation}

We update the student network with a combination of the dice loss, consistency loss and entropy loss. The optimization objection can thus be formulated as follows:

\begin{equation}
\mathcal{L}_{student}=\mathcal{L}_{dice}+\lambda\mathcal{L}_{con}+\mathcal{L}_{ent}
\label{eq7}
\end{equation}

\noindent where $\lambda$ is the hyper-parameter acting as a weight for the consistency loss. Unlike the student network, the teacher network does not participate in the back-propagation and is updated via the exponential moving average (EMA) of the weights for the current student network at each step,

\begin{equation}
\theta_{t,j}=\beta\theta_{t, j-1}+(1-\beta)\theta_{s, j}
\label{eq8}
\end{equation}

\noindent where $\theta_{t,j}$ and $\theta_{s,j}$ represent the weights for the teacher and student networks at training step $j$, respectively, and $\beta$ is a hyper-parameter for exponential moving average decay.

The ensemble of the teacher and student network makes it possible to train the network with the unlabeled COVID-19 CT images. Moreover, through the regularization of the consistency loss, the student network can learn from the teacher network output. The networks are then regularized to be transformation-consistent thus increasing the generalization capacity and robustness to the distribution difference between the lung cancer images and the unlabeled COVID-19 images.


\section{Experiments}
\subsection{Experimental settings}
\noindent \textbf{Datasets.} The lung cancer data comes from the LIDC-IDRI lung cancer dataset \cite{armato2011lung}, which is currently the largest CT dataset for pulmonary nodule detection. It provides a large number of lung CT images that share similarities with COVID-19 CT images. The LIDC-IDRI dataset contains 1018 cases, each of which includes images from a clinical thoracic CT scan along with pixel-wise annotations from experienced radiologists. The location information of every pulmonary nodule is recorded in an associated XML file for each case. The COVID-19 CT data come from a collection which is made by the Italian Society of Medical and International Radiology \cite{covid19_dataset}. This dataset includes the original CT images and COVID-19 infection masks. It contains 26 CT volumes from confirmed COVID-19 patients, and each volume contains $\sim$200 slices.

The data processing for the above two datasets is detailed as follows,

\noindent \textbf{LIDC-IDRI lung cancer dataset.} The data pre-processing steps are as follows. (1) 3D volumes to 2D slices: We reformatted all the 3D volumes into 2D slices with a size of $512\times512$. (2) Exclude scans with small nodules: To capture the presence of the nodules that might be meaningful for COVID-19 infection segmentation without decreasing the network's performance by a preponderance of smaller nodules, we excluded the scans where the pixel number for every nodule is less than 200 for robust training. (3) Generate lung nodule mask: We generated ground-truth nodule masks for scans selected by step(1) based on the patient's XML file. (4) CT image denoising: all CT scans are transformed to gray images on a Hounsfield unit (HU) scale [-600, 1500] to remove noise in CT images. The above pre-processing leaves 2,438 slices for lung cancer CT images with annotations. 

\noindent \textbf{COVID-19 dataset.} The processing of the COVID-19 dataset is similar to that of the LIDC-IDRI dataset but without step (2). After processing, there are a total of 1,616 2D COVID-19 CT slices. We employed $70\%$ of these slices as the unlabeled data for training, while the remaining $30\%$ are used to test segmentation performance. We followed the patient-level split rule when separating the COVID-19 data into a training set and a test set. Please see the detailed datasets organization in Table \ref{tab:dataset}.

\setlength{\tabcolsep}{5mm}{
\begin{table}[t]
\begin{center}
\begin{tabular}{ccc}
\hline 
 Dataset & Patients & Slices Count \\ 
 \hline 
Training set \\ (Lung cancer) & 236 & 2,438 \\ 
Training set \\ (COVID-19) & 18 & 1,156 \\
Test set \\ (COVID-19) & 8 & 460 \\ 
\hline
\end{tabular}
\end{center}
\caption{Organization of the dataset for training and test. Note that the ground-truth annotations of the COVID-19 training set are not accessible during training.}
\label{tab:dataset}
\end{table}}

\noindent \textbf{Implementation details.} 
The batch size was set to 1. We used an Adam optimizer with the initial learning rate of 6e-4 and weight decay of 0.0005. The hyper-parameters were set at $\alpha=0.005$ and $\beta=0.99$. The consistency weight was updated using a sigmoid ramp-up of $\lambda=1.5e^{-5(1-p)^2}$, where ${p}$ is the progress of the training epochs normalized to a range of 0 to 1. 

The libraries involved in this work include PyTorch 1.7.1 \cite{paszke2019pytorch} for network construction and Fourier Transform, Medicaltorch 0.2 \cite{MT-toolbox-web} for elastic transformation, Seaborn 0.11.2 for boxplot \cite{Waskom2021}, and Scikit-learn 0.24.2 \cite{scikit-learn} for feature visualization, etc. Our network is implemented on an NVIDIA RTX 2080Ti and an Intel(R) Core i7-9700K CPU.

The structure of the student and teacher networks follows U-Net architecture \cite{ronneberger2015u}, which is implemented by the built-in function of PyTorch. The trained student network is taken as the segmentation network during the test phase and outputs the segmentation results for the input COVID-19 CT images.

\noindent \textbf{Evaluation metrics.} For quantitative evaluation, we adapted the three most commonly used metrics in medical imaging analysis: the dice similarity coefficient (Dice) \cite{bertels2019optimizing}, sensitivity (Sen) \cite{zhang2019brain}, and specificity (Spe) \cite{kaushal2021firefly}. The dice similarity coefficient is an overlap index that indicates the similarity between the prediction and the ground truth. Sensitivity and specificity are two statistical measures for the performance of medical image segmentation tasks. The former measures the percentage of actual positive pixels correctly predicted to be positive, while the latter measures the proportion of actual negative pixels correctly predicted to be negative. These metrics are defined as follows:

\setlength{\tabcolsep}{1.5mm}{
\begin{table}[t]
\begin{center}
\begin{tabular}{l|ccc}
\hline
Method & Dice (\%) $\uparrow$ & Sen (\%) $\uparrow$ & Spe (\%) $\uparrow$ \\
\hline
Source-only & 40.06$\pm$8.26 & 33.68$\pm$8.55 & 99.86$\pm$0.03 \\
\hline
UA-MT \cite{yu2019uncertainty} & 40.95$\pm$5.96 & 33.40$\pm$6.49 & 99.80$\pm$0.04 \\
DAN \cite{zhang2017deep} & 42.87$\pm$3.42 & 32.81$\pm$3.81 & 99.85$\pm$0.03 \\
CTCT \cite{luo2021semi} & 44.53$\pm$6.11 & 43.73$\pm$7.05 & 99.21$\pm$0.19 \\
\hline
CCT \cite{ouali2020semi} & 45.14$\pm$4.99 & 40.35$\pm$6.89 & 99.72$\pm$0.07 \\
CPS \cite{chen2021semi} & 45.26$\pm$5.85 & 36.63$\pm$6.89 & 99.82$\pm$0.04 \\
FDA \cite{yang2020fda} & 41.55$\pm$8.75 & 38.28$\pm$7.33 & \textbf{99.96$\pm$0.01} \\
MinEnt \cite{vu2019advent} & 43.86$\pm$6.43 & 40.65$\pm$6.06 & 99.95$\pm$0.02 \\
AdvEnt \cite{vu2019advent} & 44.11$\pm$7.35 & 40.76$\pm$6.54 & 99.95$\pm$0.02 \\
\hline
MT \cite{tarvainen2017mean} & 41.49$\pm$5.28 & 30.15$\pm$5.20 & 99.90$\pm$0.02 \\
DCT \cite{qiao2018deep} & 44.42$\pm$5.34 & 34.23$\pm$5.18 & 99.85$\pm$0.04 \\
ICT \cite{verma2022interpolation} & 47.01$\pm$3.97 & 37.74$\pm$4.97 & 99.84$\pm$0.04 \\
\hline
Ours & \textbf{49.51$\pm$6.48} & \textbf{44.70$\pm$9.38} & 99.75$\pm$0.06 \\
\hline
\end{tabular}
\end{center}
\caption{Comparison between our proposed method and existing semi-supervised segmentation methods on the COVID-19 CT dataset. The highest evaluation score is marked in bold. $\uparrow$ indicates that a higher number is better.}
\label{tab:sota}
\end{table}}

\begin{equation}
\begin{aligned}
Dice = \frac{2\times{T\!P}}{2\times{T\!P}+{F\!P}+F\!N}\qquad
\end{aligned}
\label{eq9}
\end{equation}

\begin{equation}
\begin{aligned}
Sen = \frac{T\!P}{T\!P+F\!N}\qquad
\end{aligned}
\label{eq10}
\end{equation}

\begin{equation}
\begin{aligned}
Spe = \frac{T\!N}{T\!N+F\!P}\qquad
\end{aligned}
\label{eq11}
\end{equation}

\noindent where $T\!P$, $F\!P$, $T\!N$, and $F\!N$ represent the number of true positive, false positive, true negative, and false negative pixels in the prediction, respectively. 

\subsection{Experimental results}
In this section, we compared the COVID-19 infection segmentation performance of the proposed method with that of the Source-only method (U-Net trained with the lung cancer data in a supervised manner) and the state-of-the-art semi-supervised methods: UA-MT \cite{yu2019uncertainty}, DAN \cite{zhang2017deep} and CTCT \cite{luo2021semi} for medical segmentation; CCT \cite{ouali2020semi}, CPS \cite{chen2021semi}, FDA \cite{yang2020fda}, MinEnt \cite{vu2019advent}, and AdvEnt \cite{vu2019advent} for cityscapes segmentation; MT \cite{tarvainen2017mean}, DCT \cite{qiao2018deep} and ICT \cite{verma2022interpolation} originally designed for classification task, we adapted these semi-supervised classification methods to segmentation task for comparison.

\begin{figure}[!t]
\centering
\includegraphics[width=8cm]{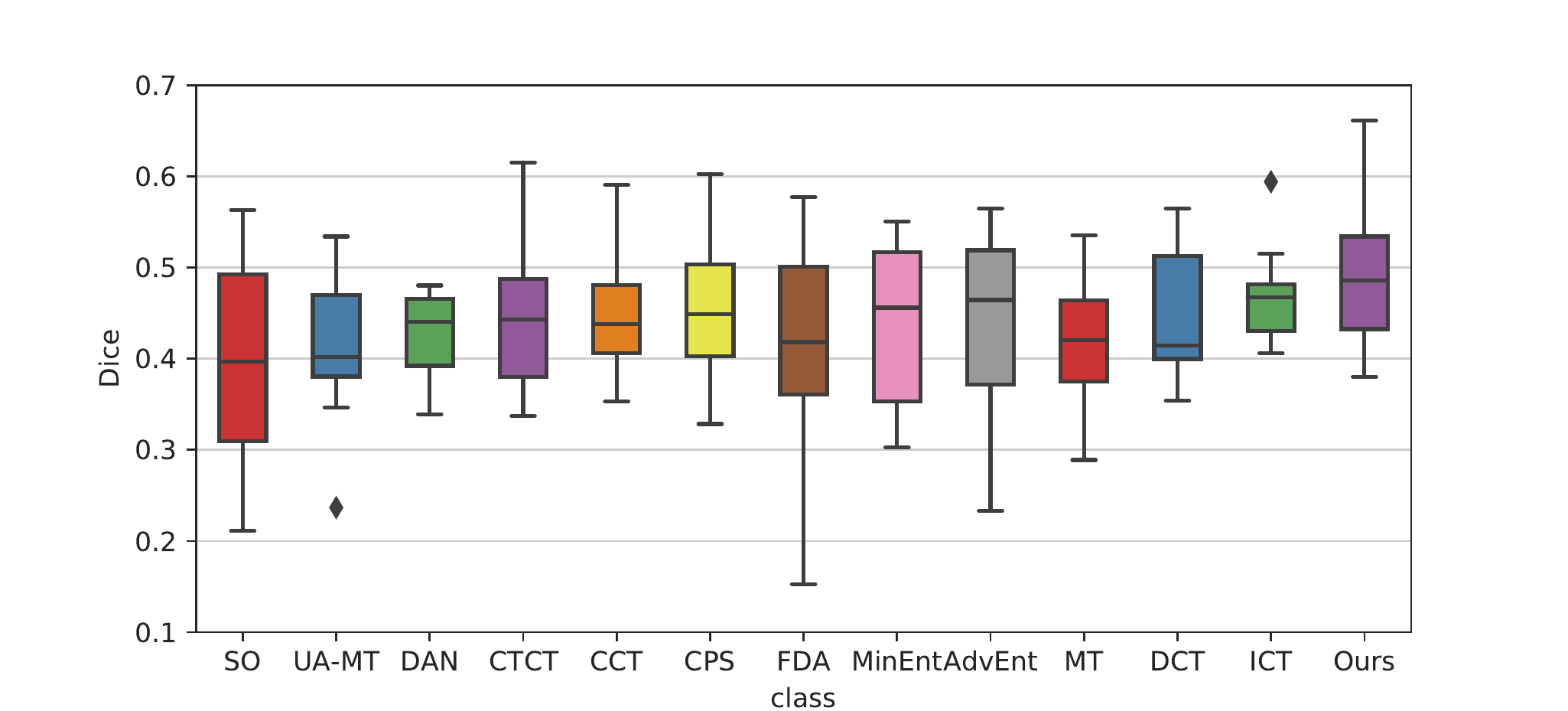}
\caption{Boxplot for the distribution of Dice similarity coefficient results on COVID-19 infection segmentation task, with the maximum/minimum results shown as upper/lower extreme, 95\% confidence interval as the color box, and outlier as dot.}
\label{fig:boxplot} 
\end{figure}


\noindent \textbf{Quantitative results.} We evaluated the segmentation performance of our proposed method through comparison with the above state-of-the-art semi-supervised methods. We also tested the robustness of the proposed method by visualizing the distribution of the results for the dice metric. For each experiment, we excluded the test set from the training of these deep learning-based models. After training, the test set was used for the segmentation test.
\begin{figure*}[!t]
\centering
\includegraphics[width=17cm, height=11cm]{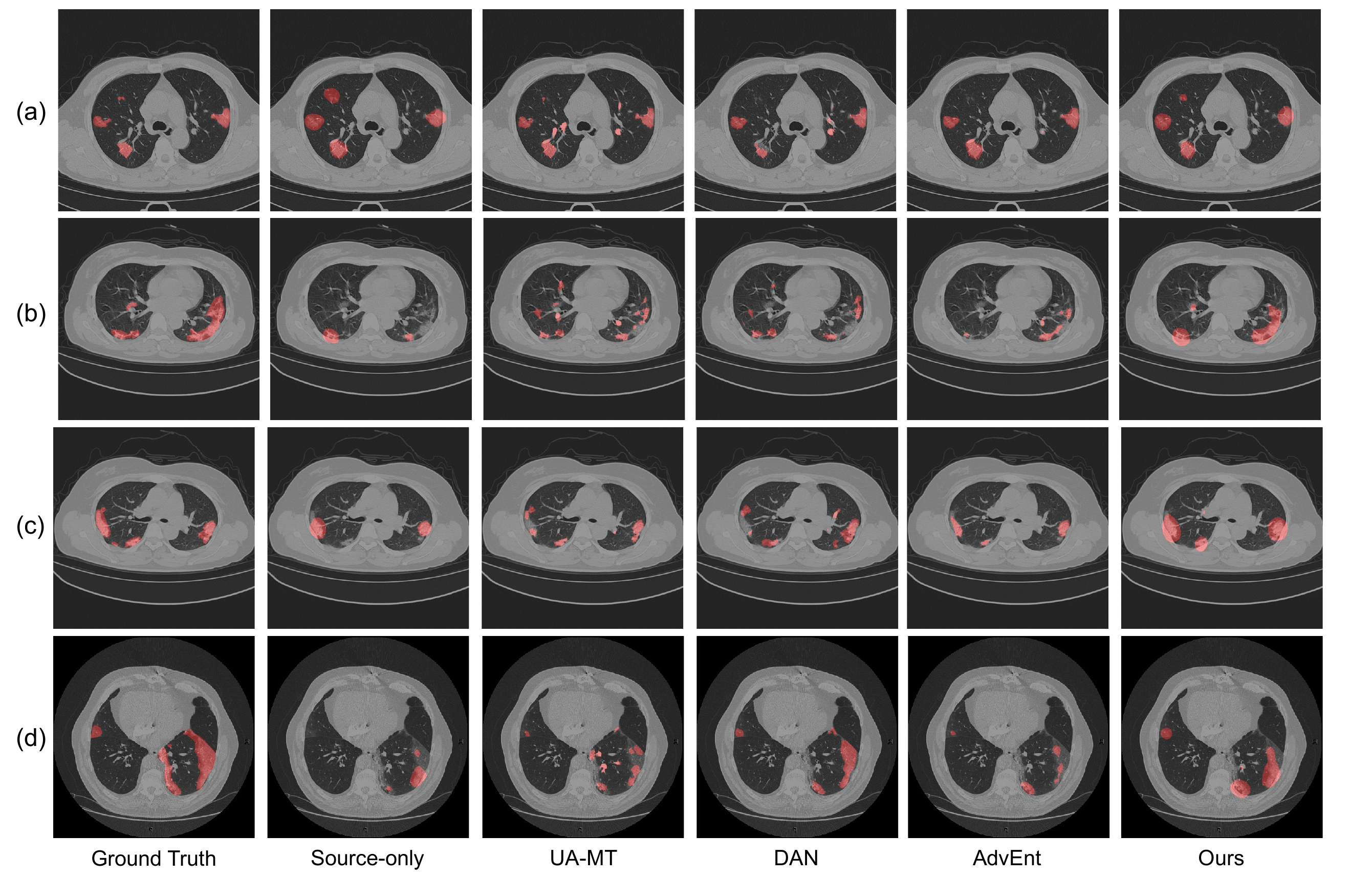}
\caption{Qualitative results for the segmentation task. The first column presents the input COVID-19 CT images with the ground truth marked in red, while columns 2 to 6 are the segmentation results of the Source-only method, UA-MT \cite{yu2019uncertainty}, DAN \cite{zhang2017deep}, AdvEnt \cite{vu2019advent}, and our proposed method.}
\label{fig:qualitative}
\end{figure*}

Table \ref{tab:sota} shows the quantitative results for each method, reported as the mean$\pm$ error interval (calculated based on a 95\% confidence interval) of the Dice similarity coefficient, sensitivity, and specificity. Generally, more accurate segmentation results are associated with higher scores for the above metrics. Our proposed method outperformed the compared methods across most metrics. Regarding dice, our method produced a 9.54\% improvement over the Source-only method, which confirms the effectiveness of our augmentation process and consistency learning. It also signifies that our approach can achieve more similarity between the predicted COVID-19 infection area and the ground truth infected area. Our method outperformed other methods on sensitivity (also known as the "true positive" rate), demonstrating that our network can identify the pixels that belong to the COVID-19 infected area more accurately. Compared with the UA-MT \cite{yu2019uncertainty}, and MT \cite{tarvainen2017mean} which also adapted teacher-student framework, our proposed method utilized elastic transformation for consistency learning, which benefits improving the robustness of our model for the shape difference between pulmonary nodule and COVID-19 infection. Our method performed competitive results on specificity (also known as the "true negative" rate), proving that our proposed model can correctly rule out pixels that don't belong to the infected area and won't generate many false-positive results.

To investigate the robustness of our proposed method, we visualized the dice similarity coefficient results for all COVID-19 test slices of our method and the compared methods by boxplot. As shown in Fig.\ref{fig:boxplot}, FDA \cite{yang2020fda} and AdvEnt \cite{vu2019advent} showed large range for extreme value, representing these methods tend to perform unstable predicted segmentation results on the different test slices. There are outliers (too low/high) appear in results of UA-MT \cite{yu2019uncertainty} and ICT \cite{verma2022interpolation}, which is also unacceptable for the real clinical application. In contrast, our proposed method yielded more stable and better results than other methods. It is mainly because the proposed CGFT-DA module can alleviate the appearance difference between lung cancer and COVID-19 images. Furthermore, our teacher-student framework enables the network to learn the robust features from unlabeled COVID-19 CT images gradually.

\begin{figure*}[!t]
\centering
\includegraphics[width=16cm]{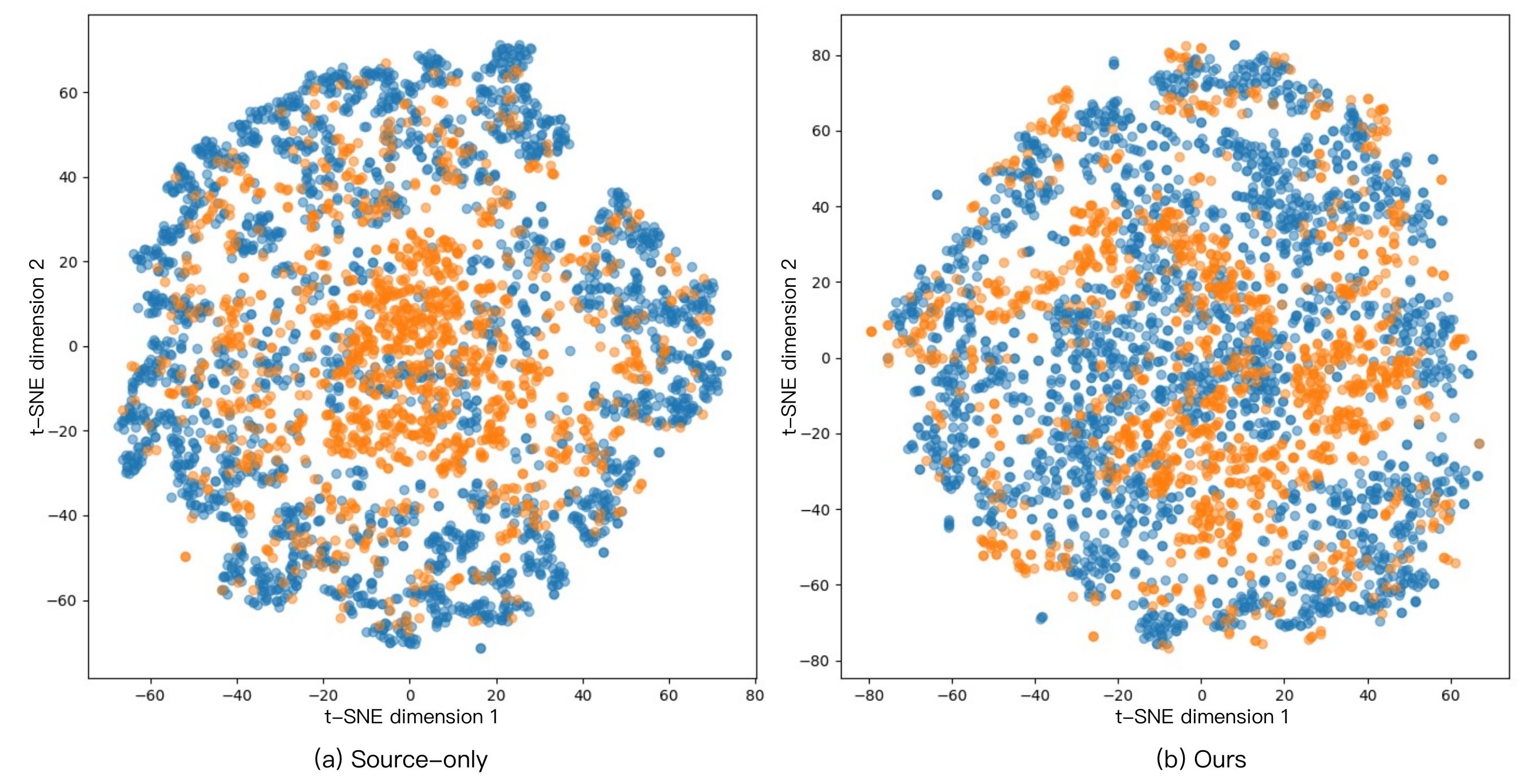}
\caption{Visualization of the feature distributions via t-SNE for two different datasets. The figure shows t-SNE visualization of the high-dimension feature extracted from (a) Source-only model trained on lung cancer dataset and (b) Our proposed model which incorporates lung cancer data and COVID-19 data for training. The blue dots correspond to the lung cancer samples, while the yellow dots correspond to the COVID-19 samples. In all cases, the Fourier Transform-based augmentation strategy and teacher-student training scheme in our method make the two feature distributions much closer.}
\label{fig:t-SNE} 
\end{figure*}



\noindent \textbf{Qualitative results.} To demonstrate the effectiveness of our proposed method from the qualitative perspective, we visualized the segmentation results of our method and some compared methods in Fig.\ref{fig:qualitative}. As shown in cases (a) to (d), the predicted infected area generated with our proposed method shows the best resemblance to the ground truth. The Source-only method can distinguish small infected regions but suffers from poor generalization for large-scale infection, such as case (b) because it is only trained with the LIDC-IDRI lung cancer dataset, in which the pulmonary nodules are relatively small and oval. UA-MT \cite{yu2019uncertainty} showed some mis-segmentation in case (a) and (b). DAN \cite{zhang2017deep} produced incomplete segmentation when handle with case (b). AdvEnt \cite{vu2019advent} also suffered from the poor generalization for the large-area COVID-19 infection in case (b) and (d). Overall, our proposed method performed better than the other methods and is consistently closer to the ground-truth COVID-19 infected region, demonstrating that our method effectively utilizes the unlabeled COVID-19 images and improves segmentation performance.

\setlength{\tabcolsep}{1mm}{
\begin{table}[!t]
\begin{center}
\begin{tabular}{l|ccc}
\hline
Method & Dice (\%) $\uparrow$ & Sen (\%) $\uparrow$ & Spe (\%) $\uparrow$ \\
\hline
Source-only & 40.06$\pm$8.26 & 33.68$\pm$8.55 & 99.86$\pm$0.03 \\
w/o CGFT-DA & 48.40$\pm$6.82 & 43.59$\pm$11.33 & 99.63$\pm$0.11 \\
w/o con. & 46.94$\pm$7.70 & 39.13$\pm$8.56 & \textbf{99.89$\pm$0.04} \\
w/o ent. & 47.98$\pm$7.07 & 44.45$\pm$9.92 & 99.72$\pm$0.05 \\
\hline
Ours & \textbf{49.51$\pm$6.48} & \textbf{44.70$\pm$9.38} & 99.75$\pm$0.06 \\
\hline
\end{tabular}
\end{center}
\caption{Ablation study of contribution of each component in our proposed network. The highest evaluation score is marked in bold. $\uparrow$ indicates that a higher number is better.}
\label{tab:ablation}
\end{table}}


\subsection{Ablation study}
This section describes the ablation study. The purpose is to assess the importance of each component in our network. Moreover, we visualized the feature distributions as learned by the proposed method and the Source-only method to prove that our proposed method could effectively align the lung cancer data and COVID-19 data in feature space.

\noindent \textbf{Contribution of each proposed component.} We validated the effect of each component of our network by analyzing the performance of the following setups: (1) Source-only, which includes only the student network and is trained with the lung cancer data in a supervised manner; (2) w/o CGFT-DA, in which the data augmentation process that is essential for overcoming the intensity difference is removed; (3) w/o con., in which the teacher network is removed, corresponding to $\lambda=0$ and the consistency loss not being used to update the student network; (4) w/o ent., in which the entropy minimization is not calculated, which corresponds to $\lambda=0$. As shown in Table \ref{tab:ablation}, the exclusion of any of the components, especially the teacher network, leads to a drop in performance, thus confirming that these components play a vital role in the performance of our proposed method. In particular, data augmentation lays a foundation for alleviating the intensity difference, while the teacher-student training scheme allows robust transformation-invariant features to be effectively exploited.


\noindent \textbf{Visualization of feature distributions via t-SNE.} To verify that our proposed method can effectively overcome the gap between lung cancer CT data and COVID-19 CT data and learn the domain-invariant features, we visualized the feature representations using t-SNE \cite{van2008visualizing}. We input the lung cancer and COVID-19 CT images to the trained Source-only model and our proposed network, respectively, then visualize the feature maps for these two data groups in low dimensions. As shown in Fig.\ref{fig:t-SNE} (a), for the Source-only model, the feature distributions of lung cancer images are separated from those of the COVID-19 images because supervision only occurs with the lung cancer data. Thus, the learned features from the Source-only cannot improve the segmentation performance for COVID-19 images. In contrast, as demonstrated in Fig.\ref{fig:t-SNE} (b), our proposed method projects the feature distributions of the two datasets into an overlapping space, illustrating the effectiveness of our scheme in aligning the two distributions. The visualization results indicate that our proposed network can learn robust domain-invariant features.

\section{Discussion}
There are several limitations of the proposed method. First, the number of patients with COVID-19 is relatively few. Since this work took place at the early stage of the outbreak of this pandemic, we could only acquire a small set of data from one institution for training our network, so the method has not shown its most potential. In the future, we will focus on training the segmentation network with more sufficient COVID-19 data, including other modalities such as CT and MRI to improve the segmentation performance further. 

Second, this work only involved the segmented COVID-19 infection in the lung. Theoretically, this method can be extended to other organs from CT images. For example, we can adapt our work for diagnosing primary hepatic tuberculosis that shares similarities with common hepatocellular carcinoma. Detecting and segmenting other organs from CT images using the proposed strategy would also be future work.

\section{Conclusion}
In this paper, we proposed a novel teacher-student-based framework for unsupervised COVID-19 infection segmentation in CT images. We attempted to address a challenging situation where there is no annotation for COVID-19 CT images, but the annotation for lung CT with pulmonary nodules is available. Given the differences between the pulmonary nodules with the COVID-19 infection, we introduced a Fourier Transform-based augmentation method to alleviate the intensity difference. We further constructed a teacher-student network that utilized the consistency loss and entropy loss to allow the proposed network to learn the robust features, thus overcoming the distribution difference to some extent. Experiments on the COVID-19 CT dataset demonstrated that the proposed method could produce a competitive performance compared with the state-of-the-art methods. In conclusion, this technique could provide accurate segmentation for treatment planning without requiring COVID-19 pixel-wise annotation acquisition, greatly facilitating the routine clinical workflow. In the future, we plan to extend our work to a multimodal framework, which learns shared representations by training with datasets from different modalities (such as MRI and CT), thereby further addressing the data shortage problem and improving segmentation quality.

\section{Acknowledgement}
This research is supported by the Government-Wide R$\&$D Fund Project for Infectious Disease Research (GFID), Republic of Korea (grant number: HG19C0682).

\bibliography{mybibfile}
\end{document}